\documentclass[twocolumn,aps]{revtex4}

\usepackage{amsfonts}
\usepackage{amsmath}
\usepackage{bm}
\usepackage{graphicx}
\usepackage{xcolor}
\usepackage{color}

\def \beq {\begin{equation}}
\def \eeq {\end{equation}}
\def \bea {\begin{eqnarray}}
\def \eea {\end{eqnarray}}
\def \bfig {\begin{figure}}
\def \efig {\end{figure}}

\def \lab {\label}

\def \bB {\bm{B}}

\def \bE {\bm{E}}
\def \bv {\bm{v}}

\def \bu {\bm{u}}

\def \bE {\bm{E}}

\def \br {\bm{r}}

\begin{document}

\title{Fluid simulations of plasma turbulence at ion scales:\\ comparison with Vlasov-Maxwell simulations}
\author{D. Perrone$^{1,2}$, T. Passot$^3$, D. Laveder$^3$, F. Valentini$^{4}$, P. L. Sulem$^3$, 
I. Zouganelis$^1$, P. Veltri$^{4}$ and S. Servidio$^{4}$}

\affiliation{
$^1$ European Space Agency, ESAC, Villanueva de la Ca\~{n}ada, E-28692 Madrid, Spain\\
$^2$ Department of Physics, Imperial College London, London SW7 2AZ, United Kingdom\\
$^3$ Universit\'e C\^ote d'Azur, CNRS, Observatoire de la C\^{o}te d'Azur, Laboratoire J. L. Lagrange, 
Boulevard de l'Observatoire, CS 34229, 06304 Nice Cedex 4, France\\
$^4$ Dipartimento di Fisica, Universit\`a della Calabria, I-87036 Cosenza, Italy}

\input epsf

\begin{abstract} 
  Comparisons are presented between a hybrid Vlasov-Maxwell (HVM)
    simulation of turbulence in a collisionless plasma 
    and fluid reductions. These include Hall-magnetohydrodynamics (HMHD) and
   Landau fluid (LF) or FLR-Landau fluid (FLR-LF) models that retain pressure anisotropy
    and low-frequency kinetic effects such as Landau damping and, for the last model, finite Larmor radius (FLR) corrections.  The problem is considered  in two space dimensions, when initial conditions involve
    moderate-amplitude perturbations of a homogeneous equilibrium plasma subject to an 
    out-of-plane  magnetic field. LF turns out to provide
    an accurate description of the velocity field up to the ion Larmor radius scale, and even to smaller scales for the magnetic field. Compressibility nevertheless appears  significantly larger at the sub-ion scales
    in the fluid models than in the HVM simulation. High frequency kinetic effects, such as cyclotron
    resonances, not retained by fluid descriptions,  could be at the origin of this discrepancy. 
    A significant temperature anisotropy is generated, with a bias towards the perpendicular component, the more intense fluctuations being rather spread out and located in a broad vicinity of current sheets.
    Non-gyrotropic pressure tensor components are measured and their fluctuations are shown to reach a significant fraction of the total pressure fluctuation, with intense regions closely correlated with current sheets. 
\end{abstract}
\date{\today} 

\maketitle 

\section{Introduction}
\label{intro}

Turbulence represents a complex and fascinating phenomenon ubiquitous in nature. It can be 
observed from simple fluids to plasmas, with compelling evidence in space plasmas observations, 
as well as in laboratory devices. The dynamics can be envisioned as a cascade of energy through 
different scales: energy is injected at large scales through inhomogeneities, and is then transferred 
self-consistently towards  smaller scales, until dissipation processes become relevant. Although in 
classical fluids this behavior is relatively well understood, in more exotic systems like plasmas, the 
cascade process is more `controversial' and, even though it has been matter of study for many 
decades, it is far from being completely understood.

In fluid systems, as described by Kolmogorov, under the assumption of isotropy and time stationarity, 
turbulence exhibits an universal behavior \cite[]{kol41}. 
Plasmas, that can be described as a fluid of particles, obey to a similar 
Kolmogorov description of the dynamics, at least
at scales large compared to the ion gyroradius. However, while energy is transferred 
to small scales, plasmas manifest their kinetic nature. With kinetic we mean here all the processes 
that cannot be described by usual collisional fluids. Characterizing the limitations of the fluid 
description of turbulence is of crucial relevance for understanding and modeling plasma dynamics.

Thanks to the contributions from space missions it is possible to benefit from a large amount 
of data to investigate the dynamics of a natural plasma, the solar wind, which can be used as a real 
plasma laboratory. \textit{In situ} observations show that the interplanetary medium is in a state of fully 
developed turbulence \cite[]{bru13}. The energy is injected at large scales, where the turbulent spectrum 
is dominated by Alfv\'enic fluctuations, characterized by a high correlation between magnetic and velocity 
fields and by incompressibility \cite[]{bel71}. In the inertial range, the turbulence spectra 
display power laws, reminiscent of the typical behavior of fluids \cite[]{mat82,tu95}. However, around ion characteristic 
scales, a break in the spectral slope is observed, with the appearance (beyond a possible 
narrow transition range \cite[]{sahraoui10}) of  steeper spectra 
extending down to the electron scales \cite[]{lea98,lea00,bal05,smi06,bou12,alexandrova12,sahraoui13,bill14} and an enhancement of magnetic 
compressive fluctuations \cite[]{ale07,ale08,ham08,sal12,kiy13,per16,per17}. In this short-wavelength 
range, often named dissipation range, plasma dynamics is sensitive to Hall dispersion and ion finite Larmor radius (FLR)
  effects, together with Landau damping and cyclotron resonances.
Moreover, spacecraft measurements reveal that the solar wind is far from the classical 
thermodynamical equilibrium, the  particle distribution functions being characterized by non-Maxwellian effects. 
In particular, ion velocity distributions show strong temperature anisotropies, with a preferential perpendicular 
heating and accelerated particles in the direction parallel to the ambient magnetic field \cite[]{mar06}. 

Plasma turbulence appears as a very complex interaction between turbulent electromagnetic fields, 
macroscopic flows, pressure anisotropy and other non-thermal effects. 
Although different physical mechanisms, such as ion-cyclotron 
resonance (e.g., \cite[]{hol02}), Landau damping 
(e.g., \cite[]{val05pre,sch09}), particle energization close to current sheets and reconnection sites 
\cite[]{ret07,sun07,osm11,cha15,cha17}, enhancement of collisions \cite[]{pezzi16,pezzi17} or stochastic heating (e.g., \cite[]{cha10}), have been proposed to explain 
\textit{in situ} observations, the role of dissipation in collisionless plasmas is still not well understood.

Nowadays numerical simulations represent an indispensable tool to 
understand the nonlinear dynamics of plasma behavior in different environments. Due to the large time 
scale separation between ion and electron dynamics, simulations that can cover the whole range of 
scales of a plasma system are very challenging. Such fully kinetic simulations
  are not really feasible with the 
available computational resources, and certain sets of approximations are needed. 

  In the study of solar wind turbulence, different numerical  models have been used, with vastly 
different assumptions and numerical schemes. The simplest fluid-like approach is given 
by magnetohydrodynamics (MHD) (e.g., \cite[]{gur05}) and all the related models that include different 
corrections and/or asymptotics, such as reduced MHD (e.g., \cite[]{zan92}),
Hall MHD (e.g., \cite[]{gho97}), electron MHD (e.g., \cite[]{kin87,gor94,cho04}), 
electron reduced MHD (e.g., \cite[]{sch09}). Other fluid-like approaches include  Landau 
  fluid (e.g., \cite[]{SHD97}) and FLR-Landau fluids (\cite[]{sul15} and references therein).
A simplified Vlasov-Maxwell approach, where the gyro-motion of the particles is averaged 
out from the system, thus restricting the dynamics to low-frequency phenomena,
is given by gyrokinetic codes (e.g., \cite[]{bri07,Howes06,Numata10}).
For a more complete kinetic description of the plasma, the most adopted numerical approach is represented 
by the Particle In Cell (PIC) methods \cite[]{bir85}. Hybrid PIC simulations (e.g., \cite[]{par09}) treat ions as 
particles and electrons as a 
neutralizing fluid, while full PIC codes (e.g., \cite[]{kar13})
describe both ions 
and electrons as kinetic particles. An alternative method to the PIC approach is provided by the Eulerian Vlasov 
model \cite[]{man02,val05,val07}. Recently, an hybrid Vlasov-Maxwell (HVM) code has been extensively 
used for the analysis of the kinetic effects in the solar wind turbulent plasma 
\cite[]{val08,val09,val10,val11,per11,ser12,per13,per14a,per14b,val14,ser14a,ser14b,vas14,val16,pezzi17a,pezzi17b}. 
The main advantage of HVM simulations is that the level of noise is negligible and the small-scale effects 
can be described with  reasonable accuracy. 

Even though
several studies have been done using various approaches, most of the results are obtained starting with 
different initial conditions or plasma parameters, thus making the comparison 
difficult. A comparative analysis of different plasma models and codes, simulating the same initial 
conditions for turbulence, and under the same plasma parameters, is  required to understand the nature
of different features recovered in turbulent systems \cite[]{hen13,par15}. 
In the present paper we focus on the properties of collisionless plasma turbulence, by a comparative 
study of three different plasma models, from fluid-like approximations to a self-consistent kinetic treatment 
of collisionless plasmas. In particular, we use (i) compressible Hall MHD (HMHD),
(ii) FLR-Landau fluid (FLR-LF), and (iii) hybrid 
Vlasov-Maxwell  (HVM) simulations, with the same geometry and initial conditions. The
  conditions of the simulations are constrained
  by the difficulty of performing kinetic simulations of developed turbulence on the present-day computers.
  We thus resorted to consider a periodic two-dimensional spatial domain and to assume relatively strong perturbations of
  a homogeneous equilibrium state in order to limit the size and duration of the simulations, although
 the considered fluid modeling of kinetic effects assumes weak perturbations.

The paper is organized as follows. In Section II, we briefly describe the three models we have considered, namely HMHD,
  LF or FLR-LF and HVM. In Section III, comparisons between the predictions of HVM with HMHD and LF are presented for a simulation with cold electrons, focusing on 
  the electromagnetic field, the ion velocity  (discriminating in particular between the solenoidal and
  compressible contribution to  the in-plane component) and, beyond the HMHD description,
  temperature anisotropy, and heat fluxes. Section \ref{sec:FLR} discusses properties of nongyrotropic pressures within HVM and also within a FLR-LF simulation with warm electrons ($\beta_e=1$).   A few concluding remarks are made in Section \ref{conclusion}.

\begin{figure}
	\begin{center}
		\includegraphics[width=0.48\textwidth]{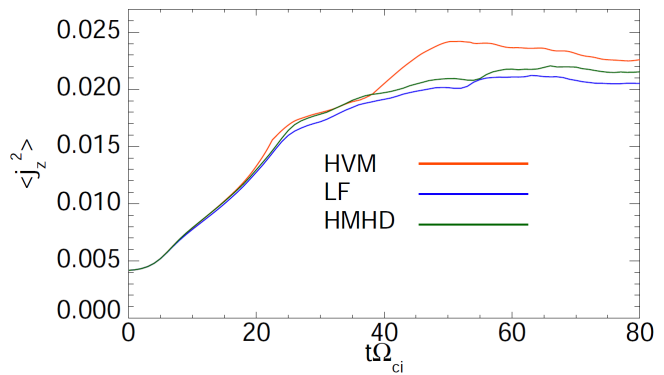} 
		\caption{Averaged out-of-plane squared current density $\langle j_z^2\rangle$ versus time for 
			HMHD (green line), LF (blue line) and HVM (red line) simulations.}
		\lab{fig:jzt}
	\end{center}
\end{figure}

\begin{figure*}
  \begin{center}
    \includegraphics[width=0.98\textwidth]{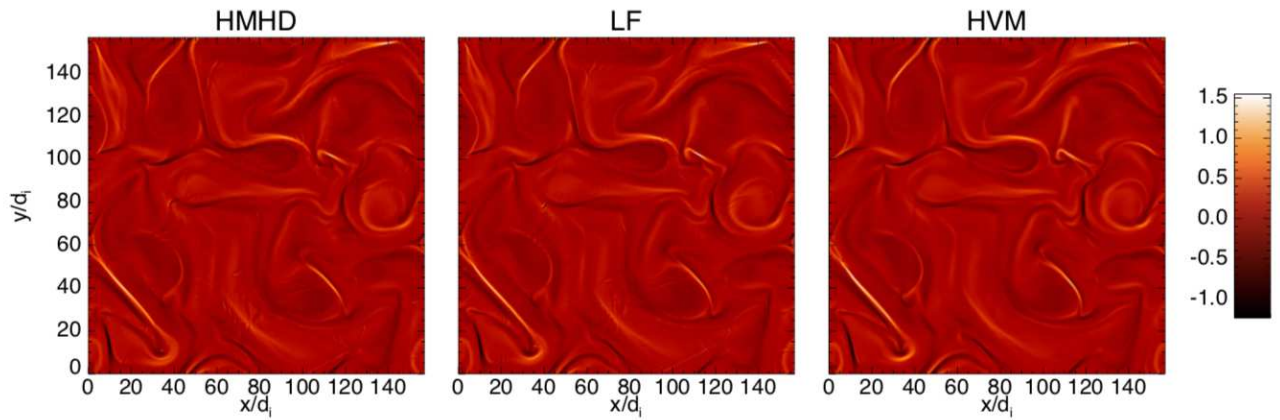}
    \caption{Color scale plot of the out-of-plane current density $j_z$ for Hall MHD (left), LF (middle) and HVM (right) simulations at $t=60$. }
    \lab{fig:jz}
  \end{center}
\end{figure*}

\begin{figure}
  \begin{center}
    \includegraphics [width=0.48\textwidth]{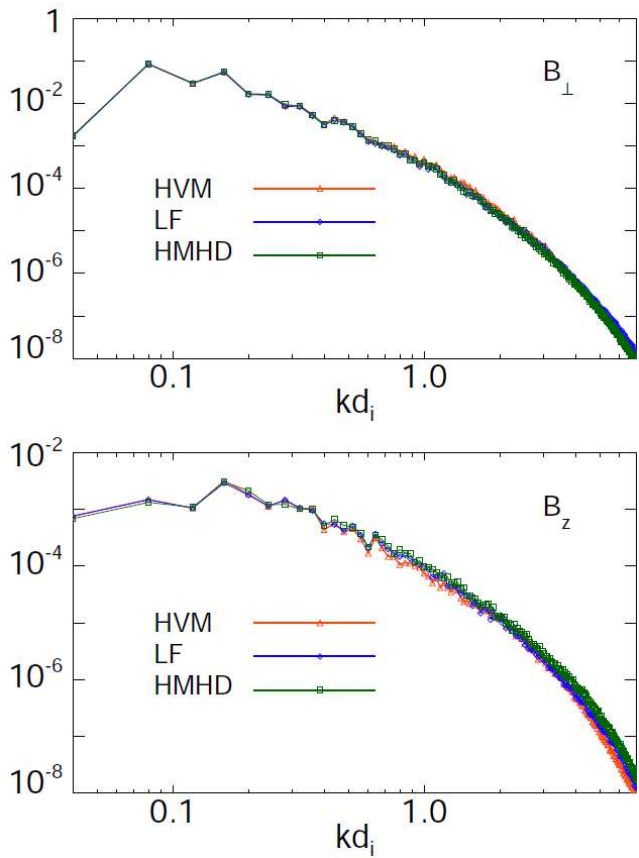}
    \caption{Spectra of the in-plane ${\mathbf B}_\perp$ (top)
      and out-of-plane $B_z$  magnetic field components, for HMHD (green solid-square line), LF 
			(blue solid-diamond line) and HVM (red solid-triangle line) simulations at $t=60$.}
    \lab{fig:spB}
  \end{center}
\end{figure}

\begin{figure*}
  \begin{center}
    \includegraphics[width=0.98\textwidth]{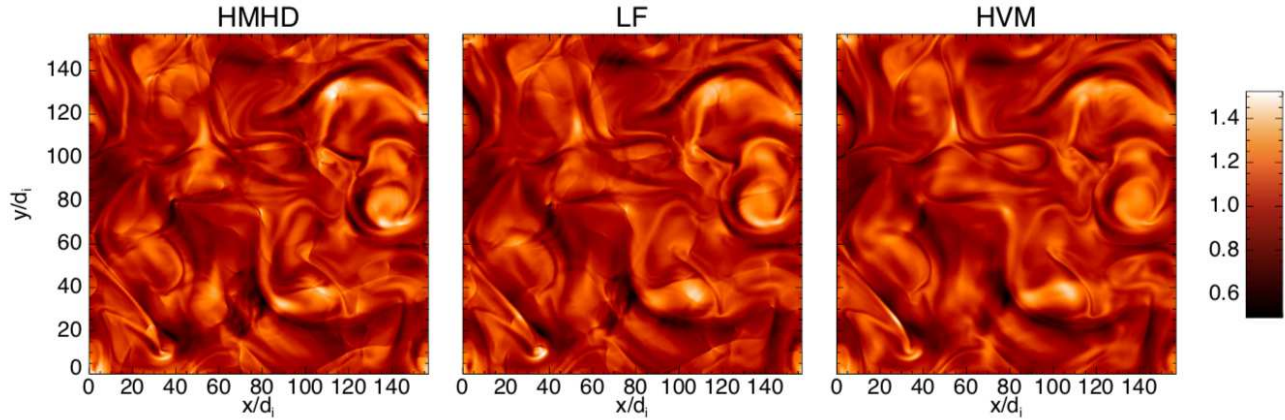}
    \caption{Color scale plot of the out-of-plane magnetic fluctuations for HMHD (left), 
      LF (middle) and HVM (right) simulations at $t=60$. }
    \lab{fig:bz}
  \end{center}
\end{figure*}

\begin{figure}
  \begin{center}
    \includegraphics [width=0.48\textwidth]{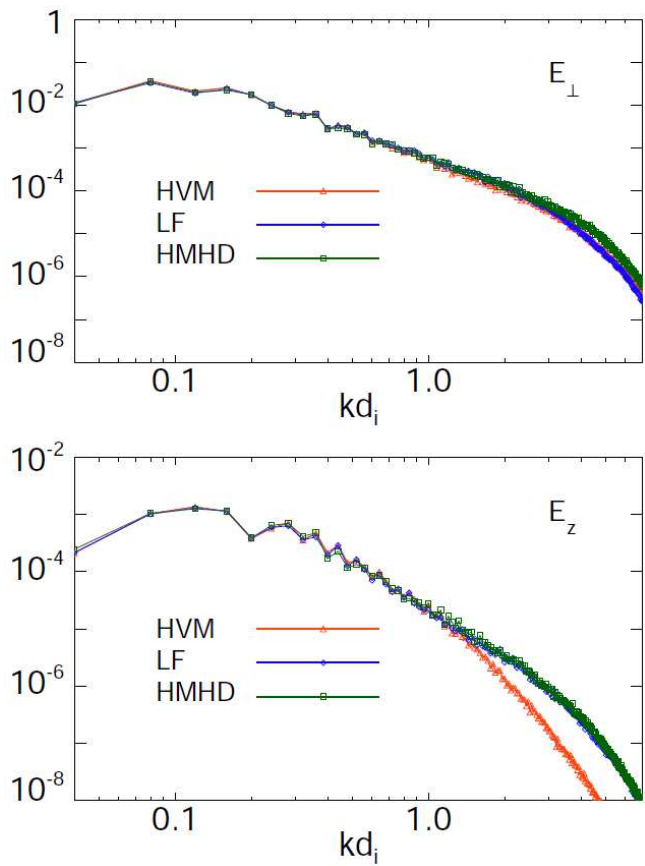}
    \caption{Spectra of in plane and out-of-plane plasma electric field for HMHD (green solid-square line), LF 
			(blue solid-diamond line) and HVM (red solid-triangle line) simulations at $t=60$.}
    \lab{fig:spe}
  \end{center}
\end{figure}

\begin{figure*}
  \begin{center}
    \includegraphics[width=0.98\textwidth]{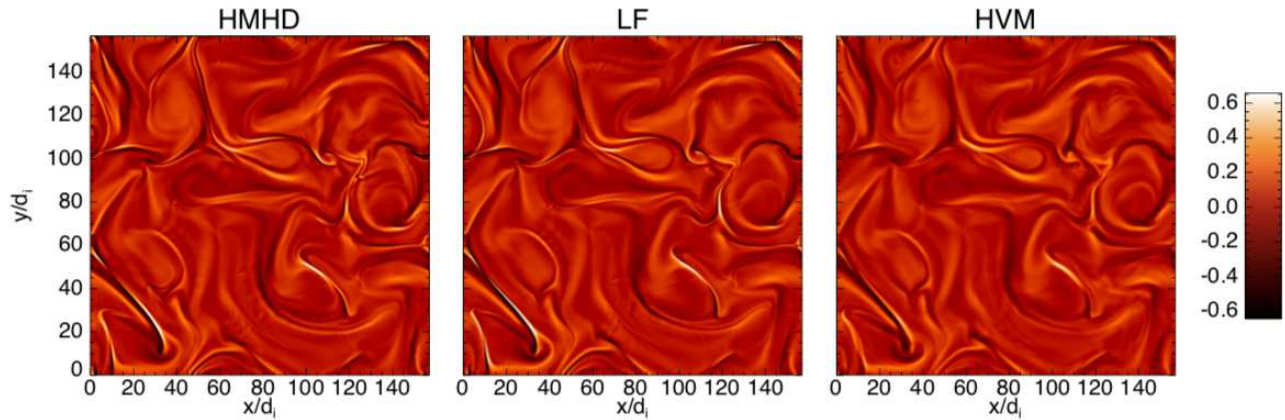}
    \caption{Color scale plot of the out-of-plane vorticity $\omega_z$ for HMHD (left), 
      LF (middle) and HVM (right) simulations at $t=60$.}
    \lab{fig:omegaz}
  \end{center}
\end{figure*}

\begin{figure}
  \begin{center}
    \includegraphics [width=0.48\textwidth]{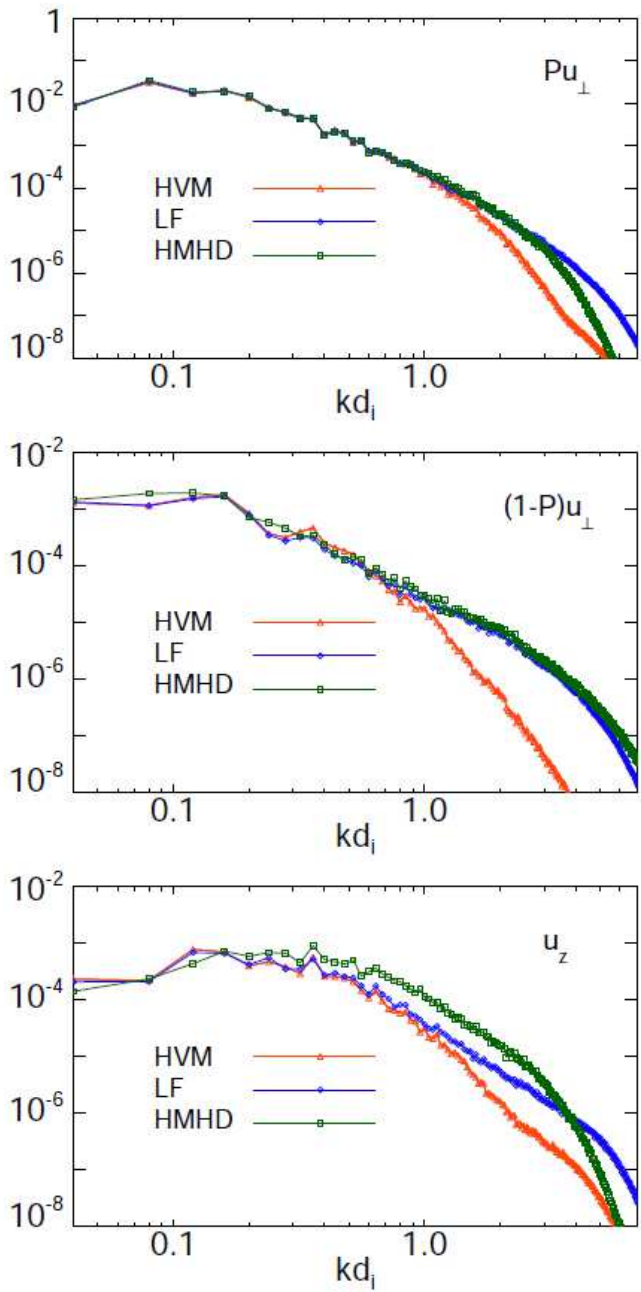}
    \caption{Spectra of the solenoidal  $P{\mathbf u}_\perp$ (top)
      and compressible $(1-P){\mathbf u}_\perp$  (middle) components of the
      in-plane velocity, and of  out-of-plane velocity $u_z$ (bottom)
      for HMHD (green solid-square line), LF 
			(blue solid-diamond line) and HVM (red solid-triangle line) simulations at $t=60$.}
    \lab{fig:spu}
  \end{center}
\end{figure}

\begin{figure*}
  \begin{center}
    \includegraphics[width=0.98\textwidth]{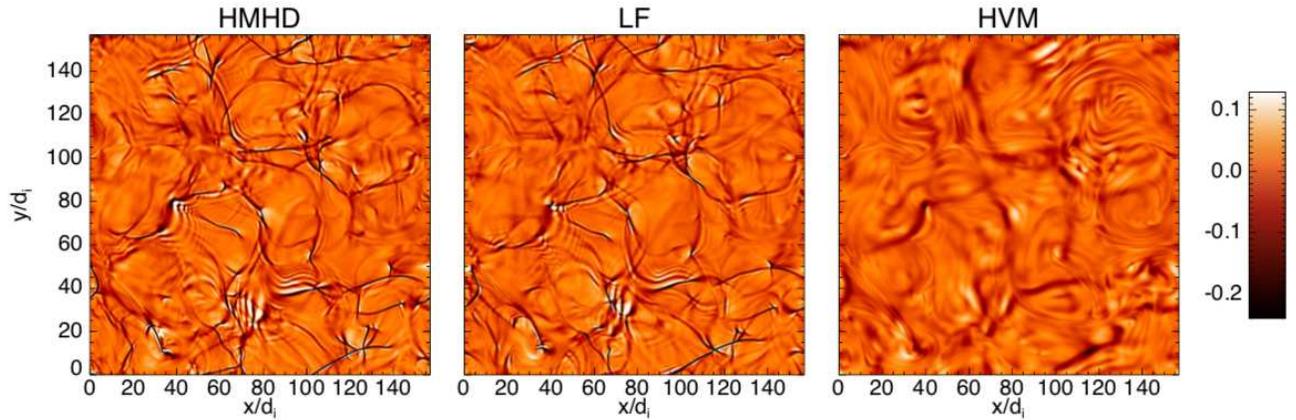}    
    \caption{Color scale plot of the ion velocity divergence $\xi={\bf \nabla}\cdot{\mathbf u}$
      for HMHD (left), LF fluid (middle) and HMV (right) simulations at $t=60$.}
    \lab{fig:divv}
  \end{center}
\end{figure*}

\begin{figure}
  \begin{center}
    \includegraphics [width=0.48\textwidth]{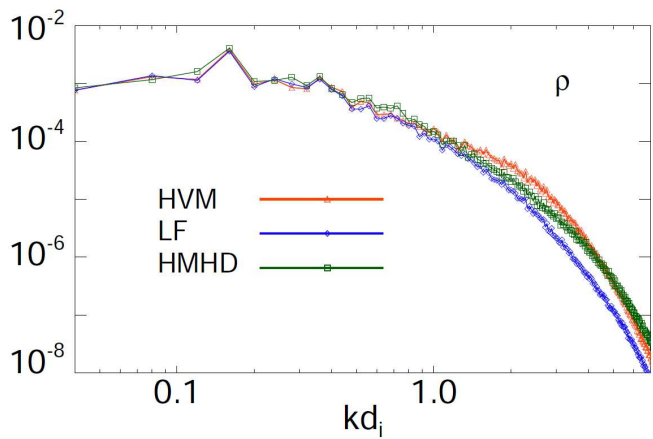}   
    \caption{Spectra of plasma density $\rho$ for HMHD (green solid-square line), LF 
			(blue solid-diamond line) and HVM (red solid-triangle line) simulations at $t=60$.}
    \lab{fig:sprho}
  \end{center}
\end{figure}

\begin{figure}
  \begin{center}
    \includegraphics [width=0.48\textwidth]{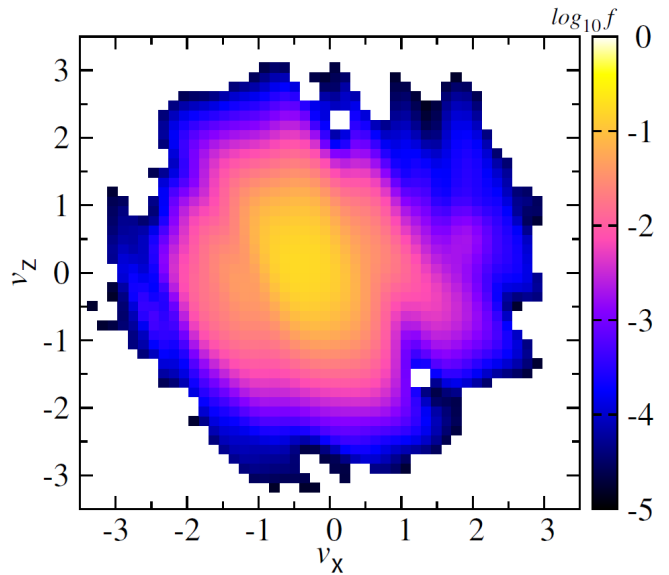}
    \caption{Color scale plot of the proton velocity distribution in the $v_x$--$v_z$ plane at the
      end of HVM simulation.}
    \lab{fig:distrib}
  \end{center}
\end{figure}

\begin{figure*}
  \begin{center}
    \includegraphics [width=0.98\textwidth]{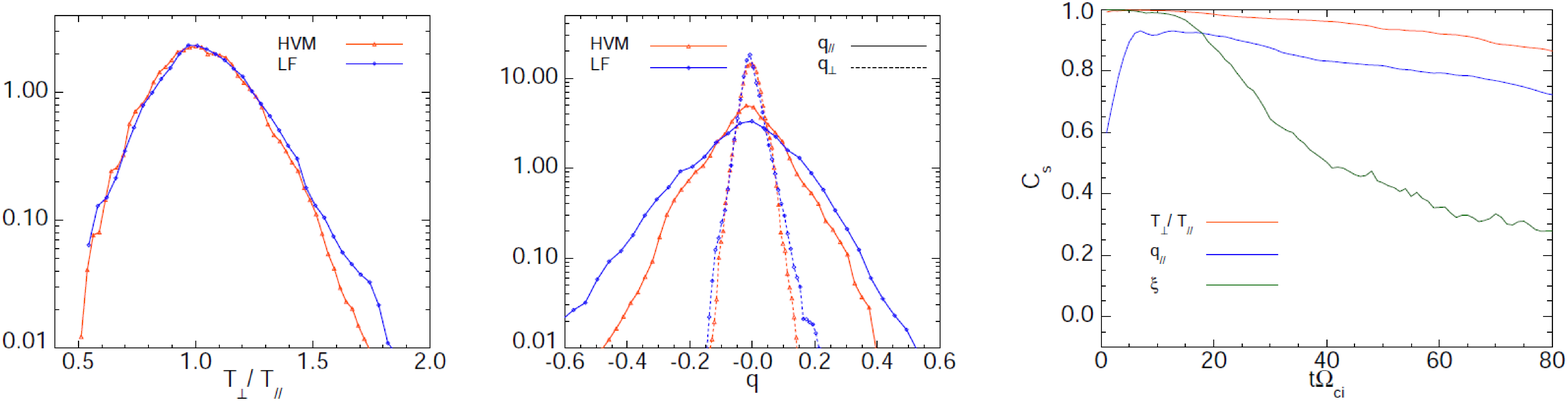}
    \caption{PDFs of the ion temperature anisotropy, $T_{\perp}/T_{\parallel}$ (left), 
      parallel, ${q}_{\parallel}$ (solid lines), and perpendicular, ${q}_{\perp}$ (dashed lines), 
      heat flux in the reference frame of the local magnetic field (middle) 
      for LF (blue-diamond line) and HVM (red-triangle line) simulations at $t=60$. Right panel shows the time evolution of the
      the Spearman correlation coefficient between LF and HVM results for the proton temperature anisotropy (red line),
      parallel heat flux (blue line) and compressibility (green line). }
    \lab{fig:aniso}
  \end{center}
\end{figure*}

\begin{figure}
  \begin{center}
    \includegraphics [width=0.48\textwidth]{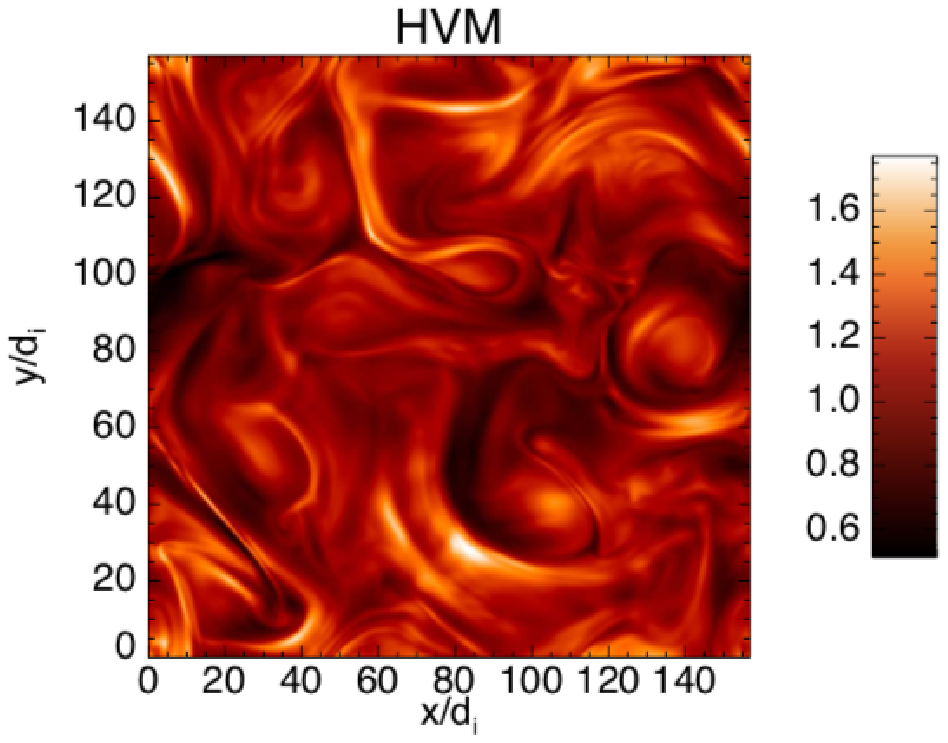}
  \end{center}
  \caption{Color scale plot of the temperature anisotropy $T_\perp/T_\|$ for the HVM run at $t=60$.}
  \lab{fig:tperpovertpar}
\end{figure}

\section{The models}
\label{models}

The dynamical behavior of  plasma fluctuations depends markedly on their frequencies. At the lowest frequencies, ions and electrons are locked together by electrostatic forces and the plasma behaves like an electrically conducting fluid; this is the MHD.  Note that, when collisions are sufficiently weak, pressure anisotropy can also develop. At somewhat higher frequencies,
electrons and ions can move relatively to each other, behaving like two separate and 
interpenetrating fluids; this is the two-fluid regime. At still higher frequencies, a  kinetic theory, where the statistical description of the plasma state is provided by the particle distribution functions in phase space, is required.

In this section, we provide a description of the models used in the present paper. The simplest 
treatment is given  by HMHD that retains dispersive effects  
which become  significant at scales comparable to the ion skin depth $d_i$. 
Then, we focus on FLR-LF,  that
also retains pressure anisotropy and low-frequency
kinetic effects such as Landau damping and finite Larmor radius corrections at scale comparable to or
smaller than the ion gyroradius. When the latter effect is neglected, we refer to the model
as LF. Finally, we describe the HVM approach, where 
the evolution of the ions is obtained by solving an equation for the ion distribution function.
Electron inertial effects have been included in all three cases with a mass ratio $m_i/m_e=100$.  
Both in fluid and HVM simulations, electrons are treated as a cold fluid. The case of
a FLR-LF simulation with a warm isothermal electron fluid with $\beta_e=1$ is also considered since, 
as discussed below, the cold electron FLR-LF simulation can only be integrated up to an intermediate time 
after which numerical difficulties in the calculations of the FLR terms are encountered. 
The results of Section \ref{results} thus involve a precise comparison between HMHD, HVM and LF simulations. 
The discussion of FLR corrections in the warm case is presented in Section \ref{sec:FLR}.

In both fluid and kinetic simulations, at t=0, the equilibrium consists of a homogeneous
plasma embedded in a uniform background magnetic field, $\bB_0$, along the $z$-direction. 
In the HVM run, ion distribution function is initialized with
a Maxwellian with homogeneous density. The system evolution is investigated in a double periodic 
domain $(x,y)$ perpendicular to $\bB_0$.
The equilibrium configuration is initially perturbed by velocity and magnetic field  fluctuations 
with wavenumbers in the 
range $0.08 < k < 0.24$ in units of $d_i^{-1}$, where $k = 2 \pi m/L$, with $2 \leq m \leq 6$
(here $L$ being the box size in each spatial direction). The phases of these perturbations are random.
The rms of the magnetic perturbations is $\delta b/B_0 \sim 0.3$ and neither density 
disturbances nor parallel variances are imposed at $t=0$. The ion plasma beta is $\beta_i = 1$, defined
as $2v_{th,i}^2/v_A^2$, where $v_{th,i} = \sqrt{T_i/m_i}$ is the ion thermal speed
and $v_A = B_0/\sqrt{4\pi n_0 m_i}$ the Alfv\'en speed.
The system size in the spatial domain 
is $L = 2\pi \times 25d_i$ in both $x$ and $y$ directions, discretized with $1024^2$ grid-points in the 2D spatial domain. For the HVM run, the 3D velocity domain, limited by  $\pm 5 v_{th,i}$ 
in each directions, is discretized with an uniform grid of $51^3$ points.

\subsection{Hall-MHD}
\label{HMHD}

The compressible HMHD equations including a modeling of electron inertia \cite{val07} read
\bea
\label{eq:HMHD1}
\frac{\partial \rho}{\partial t} + \nabla \cdot \left(  \rho \bu \right) = 0, \\
\label{eq:HMHD2}
\frac{\partial \bu}{\partial t} + \left( \bu \cdot \nabla \right) \bu = 
-\frac{ {\beta}}{2\rho} \nabla p  + \frac{1}{\rho} 
\left[  \left(  \nabla \times \bB \right) \times \bB \right], \\
\label{eq:faraday}\frac{\partial \bB}{\partial t} = - \nabla \times \bE+\eta\nabla^2\bB,\\
\label{eq:HMHD3}
(1+\alpha\nabla^2) \bE + \left[  \bu \times \bB - 
\frac{1}{\rho}\left(  \nabla \times \bB \right) \times \bB \right]=0\\
\label{eq:HMHD4}
p = \rho^\gamma,
\eea
where $\rho$ is the plasma density (only due to ions), $\bu$ the hydrodynamic velocity of the plasma, $\bB$  the magnetic field, 
$p$ the total pressure (taken isotropic), $\gamma = 5/3$   the 
adiabatic index, $\eta=0.02$ the magnetic diffusivity and $\alpha=1/100$ an artificial electron to ion mass ratio. 
Equations (\ref{eq:HMHD1})-(\ref{eq:HMHD4}) are written in a dimensionless form. 
In particular, $\rho$ is normalized to $n_0 m_i$; $\bu$, to the Alfv\'en speed; 
$t$, to the inverse ion-cyclotron frequency $\Omega_{ci}^{-1}$ and the unit length to
the ion skin depth $d_i$. 

All the fluid numerical codes employ a 2D Fourier pseudo-spectral method to calculate spatial derivatives, 
and time integration is performed via a third-order Runge-Kutta scheme. Aliasing errors in the evaluation 
of nonlinear terms are partially removed by a 2/3 truncation in the spectral space.
 
\subsection{Landau fluid and FLR-Landau fluid}
\label{LF}

The FLR-Landau fluid model appears as a system of dynamical equations for the magnetic field
  and for the density, velocity,  parallel and perpendicular pressures and heat fluxes of the ions.
The electric field is given by a generalized Ohm's law which includes the Hall term, 
the electron pressure gradient (when present), together with electron inertia with the same simplified form as in the HMHD case. The fluid hierarchy is closed by expressing the gyrotropic fourth-rank moments 
in terms of the above quantities, in a way consistent with 
the low-frequency linear kinetic theory. Two cases will be considered. In the simple LF case, ion linear Landau damping only is retained while the FLR-LF case also
takes into account FLR corrections. These corrections are evaluated algebraically
in terms of the retained quantities. The equations are too 
complicated to be explicitly written here, but can be found in \cite{sul15}, namely in Section 2 (where the terms 
corresponding to the work of the nongyrotropic pressure force are neglected), in Eqs. (3.22) and (3.23) 
and in Subsections 5.1.1-5.1.4. 
A main assumption for modeling Landau damping (and thus closing the fluid hierarchy)
is that, up to the distortion of the magnetic field lines, 
Landau damping keeps the same form as in the linear regime, an hypothesis which requires that
the turbulence fluctuations are not too strong (a condition which is unfortunately not met in the  simulations presented in this paper). This modeling involves a Hilbert transform along the distorted
magnetic field lines which leads to Landau dissipation. Its approximation in the numerical code is discussed in
\cite{pas14}. FLR-LF accuracy was successfully tested in the linear regime  in three space dimensions, in particular
at the level of the dispersion relation, the damping and the polarization of kinetic Alfv\'en waves
(e.g., \cite{hun13}). Three-dimensional simulations of kinetic Alfv\'en wave turbulence using this model  are presented in \cite{koba17}.
In the present two-dimensional regime, the situation is more delicate, as the ambient field does not
affect the transverse dynamics,
making Landau damping only due to the transverse distortion of the magnetic field lines.

\subsection{Hybrid Vlasov-Maxwell}
\label{HVM}

We solve the Vlasov-Maxwell equations in the hybrid approximation (kinetic ions and
fluid electrons) by using
the HVM code \cite[]{val07}, in a 2D3V phase space domain (two dimensions in 
physical space and three dimensions in velocity space). The Vlasov equation is integrated for the ion 
distribution function, while the electron response is taken into account through a generalized Ohm's law 
for the electric field. The dimensionless HVM equations are given by
\bea
\label{eq:vlas}
\frac{\partial f}{\partial t} + \bv \cdot \frac{\partial f}{\partial \br}+\left( \bE + \bv \times \bB \right) 
\cdot \frac{\partial f}{\partial \bv} = 0,
\eea
supplemented by Eqs. (\ref{eq:faraday})-(\ref{eq:HMHD3}), where $f(\br,\bv,t)$ is the ion distribution function. The ion density, $n$, 
and the bulk velocity, $\bu$, are evaluated as velocity moments of the ion velocity distribution. 
The quasi-neutrality approximation, $n = n_i =n_e$, is used and cold electrons are considered. 
As in the previous models, time is scaled by the inverse ion-cyclotron frequency, $\Omega_{ci}^{-1}$; velocity 
by $v_A$; lengths by the ion skin depth, $d_i = v_A/\Omega_{ci}$; and masses by the ion mass, 
$m_i$. From now on, all physical quantities will be expressed in units of the characteristic 
parameters listed above.  

The numerical solution of the Vlasov equation is based on the well-known time splitting method,
 first proposed in \cite[]{che76}. The time splitting consists in separating the evolution of the particle 
 distribution function in phase space into subsequent translations, first in physical space and then in 
 velocity space \cite[]{man02}.  Finally, the Current Advance Method (CAM) \cite[]{mat94,val07} provides numerical 
solutions for the time advancement of electric and magnetic fields. The time step, $\Delta t$, 
has been chosen in such a way that the Courant-Friedrichs-Lewy condition for the numerical stability 
of the Vlasov algorithm is satisfied \cite[]{pey86}. { It is to be noted that the monocore computer time per time step on machines of comparable speed is roughly 450 time larger for the Vlasov simulation than for the FLR-LF run.}

\section{Comparisons of HMHD, LF and HVM simulations }
\label{results}

Large-scale fluctuations produce a turbulent cascade toward kinetic scales, as in the fluid counterpart.  
In analogy with fluid models for plasma in decaying turbulence \cite[]{min09}, it exists an instant of time, $t^*$, 
at which the turbulence reaches its maximum activity. Near this particular time, decaying turbulence shares 
many statistical similarities with steady state (driven) turbulence. It is possible to identify $t^*$ by following 
the temporal evolution of the averaged out-of-plane squared current density $\langle j_z^2\rangle$, which is
a good indicator of the level of the turbulent activity. It is important to note that the
  maximum value of $\langle j_z^2\rangle $ is prescribed by the strength of the dissipation processes. All simulations
  include a magnetic diffusivity term with a coefficient $\eta =2\times \cdot 10^{-2}$. In the fluid runs, additional regularizing terms  are to be supplemented
  to smooth the equations and produce a similar behavior of $\langle j_z^2\rangle$. For
  this purpose, we add a hyperviscosity and an hyperdiffusivity, as bi-Laplacians, with coefficients
  $\nu_4= \eta_4=5\times 10^{-4}$ in the case of HMHD and $ \nu_4 = \eta_4 = 10 ^{-4}$ for  LF 
  In the latter model, bi-Laplacian dissipative terms are also added, 
  with a coefficient equal to $2.5\times 10^{-3}$  in the equations for the density and the pressures, and   $10^{-4}$ in the equations for the heat fluxes in order to deal with the high level of compressibility in the simulation (the Mach number reaching values up to $0.4$). { Note that all the 
    above dissipative terms have been added only for numerical reasons, in order to    
    prevent spurious effects such as artificial filamentation of current sheets and 
    shock-like structures. However, we verified that these non-ideal contributions are 
    only important at wavenumbers $k$ such that at $k d_i > 3$, thus at scales significant 
    smaller than the turbulent inertial range. This high-precision tuning of the artificial 
    dissipation terms, achieved through several convergence tests, is particularly 
    important for fluid models that, even in their extensions retaining Landau damping, do 
    not capture all the dissipative processes acting at the kinetic scales.}
    
Figure~\ref{fig:jzt} shows the behavior of $\langle j_z^2\rangle$ as a function of
time for the corresponding HMHD (green line), LF (blue line) and HVM (red line) simulations. This color code will be used in all
the other graphs presented in this paper. The similarity between the turbulence activity
in the different models permits comparisons between these descriptions. Note also that the same $\beta$ parameter was used for the three simulations, thus ensuring  a similar level of density fluctuations (as can be checked below by inspection of the density spectrum). Whereas there is no ambiguity in the choice of $\beta$  for the LF and HVM runs, since they both share the same linear theory, for HMHD no proper choice of $\beta$ would however permit to recover the same linear theory. Consequences will be discussed below at the level of the velocity field divergence.

\subsection{Turbulence and coherent structures}
\label{turbulence}

\begin{table}[htp]
\begin{center}
\begin{tabular}{|c||c|c|c|}
 \hline
 \null    & {\bf HMHD}/{\bf LF} & {\bf HMHD}/{\bf HVM} & {\bf LF}/{\bf HVM} \\
 \hline
 $b_x$ &   99\% &  99\%   & 99\% \\
 \hline
 $b_y$ &   99\% &  99\%   & 99\% \\
 \hline
 $b_z$ &   82\% &  80\%   & 93\% \\
 \hline
 $j_z$  &   90\%  &  85\%   & 91\% \\
 \hline
 $e_x$ &   94\% &  94\%   & 97\% \\
 \hline
 $e_y$ &   94\% &  92\%   & 96\% \\
 \hline
 $e_z$ &   89\% &  90\%   & 96\% \\
 \hline
 $\rho$ &   85\%  & 80\%  & 92 \% \\
 \hline
 $u_x$ &   95\% &  95\%   & 98\% \\
 \hline
 $u_y$ &   96\% &  96\%   & 99\% \\
 \hline
 $u_z$ &   67\% &  49\%   & 83\% \\
 \hline
 $\omega_z$  &   86\%  &  82\%   & 90\% \\
 \hline
 $\xi$  &  9\%   & 31 \%  &   35\% \\
 \hline
\end{tabular}
\end{center}
\caption{Spearman correlation coefficients between several quantities 
of the three runs, i.e. HMHD/LF (left column), HMHD/HVM (center column) 
and LF/HVM (right column), at $t=60$.} \label{tb:spearman}
\end{table}
\begin{table}[htp]
	\begin{center}
		
		\begin{tabular}{|c||c|}
			\hline
			\null    & {\bf LF}/ {\bf HVM} \\
			\hline
			$p_\perp $ &92 \% \\
			\hline
			$p_\| $ &89 \% \\
			\hline
			$T_\perp$  &91 \% \\
			\hline
			$T_\|$ &88 \% \\
			\hline
			$T_\perp/T_\|$  &92 \% \\
			\hline
			$q_\perp$ & 37 \% \\
			\hline
			$q_\|$ & 79 \% \\
			\hline
		\end{tabular}
		
	\end{center}
	
	\caption{Spearman correlation coefficients between the kinetic quantities 
		of the LF and HVM simulations at $t=60$.}
	\label{tb:spearman_kinetic}
\end{table}

Turbulence leads to the generation of coherent structures, such as current and vorticity
sheets. Figure~\ref{fig:jz} displays,  for the three models, a  color plot of the 
out-of-plane current density, $j_z = (\nabla \times {\bf B})_z$ at time $t=60$, during the period of maximal turbulence activity.
Note that in this figure and in all the other color scale plots presented in this paper, the extremal values of the fields can exceed
the limits of the color bar ranges in some very localized regions. This choice was made in order to improve the visibility of the
structures of intermediate amplitude.
Comparison between HMHD (left panel), LF (center panel) and HVM (right panel) simulations 
shows that there are strong similarities regarding 
the in-plane magnetic field. Regions of very intense $j_z$ are in particular present in each panels 
at the same locations. 

In order  to get a more quantitative assessment of the similarity between the three runs in physical space,
we evaluate at $t=60$ the Spearman correlation coefficients, defined as
\beq
C_s(f,g) =  \frac{\sum_i(Rf_i-\bar{f})(Rg_i-\bar{g})}{\sqrt{\sum_i(f_i-\bar{f})^2}\sqrt{\sum_i(g_i-\bar{g})^2}} \ 
\eeq 
where $f_i$ and $g_i$ are the values at the grid points of the  same scalar field provided by different models
at a given time, 
$R$ is an operator ranking these values in a decreasing order, and $\bar{f}$ and $\bar{g}$ 
denote the mean values of the corresponding fields at the considered time. 
An overview of the values of correlation $C_s$ for various fields
between the three runs, i.e. HMHD/LF, HMHD/HVM and LF/HVM, at $t=60$,
is reported in Table \ref{tb:spearman}. More specifically, we obtained for the out-of-plane current
a correlation coefficient of $85\%$ for
HMHD/HVM and $91\%$ for LF/HVM. This result indicates a high correlation at
the level of the in-plane magnetic field. For $B_x$ and $B_y$,  the correlation is indeed of $99\%$ both HMHD/HVM and LF/HVM.

For a  Fourier space characterization of the in-plane magnetic field in the three models,
  we plot in Fig. \ref{fig:spB} (top panel)
  the energy spectra of the in-plane components ${\mathbf B}_\perp$
  of the magnetic field fluctuations. We observe that these spectra are very close
  for the three models up to $kd_i=5$.

In order to analyze the out-of-plane magnetic fluctuations $B_z$,
  we display in Fig. \ref{fig:bz} the color plot
  of this field in the three models. In spite of a global similarity, discrepancies are visible at small scales. A conspicuous difference concerns the presence of steepened waves-like features (probably associated with fast waves) in the fluid runs, which are absent (or at least heavily damped) in the Vlasov case.
  The coefficient $C_s$ is indeed $80\%$ for HMHD/HVM and $93\%$ for LF/HVM. The difference between the
  three runs are also visible on the energy spectrum shown in Fig. \ref{fig:spB} (bottom panel),
 for wavenumber close to  $kd_i=1$, LF being slightly closer to HVM than HMHD is.

It is also of interest to compare the electric field given  by the various descriptions,
  as it is commonly measured by satellite missions and is at the origin of particle acceleration,
  although in the numerical simulations
  it is constructed from various other fields by means of the generalized Ohm's law. Analysis in physical space   shows that all three components display  correlation coefficients larger than $90\%$ and  $96\%$ for HMHD/HVM and LF/HVM respectively.
  The spectrum of the in-plane electric field  displayed in Fig. \ref{fig:spe} are similar throughout the whole spectrum for the three descriptions, the agreement LF/HVM being slightly better than HMHD/HVM. For the out of plane component, both fluid descriptions disagree with HVM result for $kd_i>1$. 

Turning to the velocity field, we first plot in Fig. \ref{fig:omegaz} the out-of-plane vorticity
  $\omega_z$ at  $t=60$ obtained
  with the three models. A good  similarity is observed, supported by the results of correlation measurements (see Table \ref{tb:spearman}), implying  a good agreement  at the level
  of the (dominant) solenoidal part of the in-plane velocity. In Fourier space, the energy spectrum of this field,
  displayed in Fig \ref{fig:spu} (top panel), reveals a satisfactory match between both fluid models and  HVM up to $kd_i= 1$.
  
Although significantly smaller than the solenoidal part (a factor $4$ in the rms values), the compressible component of
  the in-plane velocity is very   different in the three simulations, as visible on the corresponding
  color plots of $\xi=\nabla \cdot {\mathbf u}$ in Fig.~{\ref{fig:divv}}. Both for HMHD/HVM and LF/HVM, the correlation coefficient of $\xi$ is of order $30\%$. The HVM large-scale behavior of $\xi$ is relatively well reproduced with the fluid simulations. On the other hand, the small-scales in HVM are smoothed out, thus explaining this intermediate value of the correlation. The poor agreement between HMHD and LF is rather due to the fact that the small-scale features of the fluid simulations are completely out of phase due to a difference in the description of the fast waves linear behavior, as mentioned in Sec. \ref{models}.
Shock-like cellularization, typical of compressible turbulence and  evidently due to an higher
in-plane magnetosonic activity, is quite clear in HMHD and LF models. The fronts are significantly
  mollified in HVM. This difference suggests that 
most likely, in a fully kinetic treatment, compressible activity is balanced by 
Landau damping (approximately modeled in LF) and ion cyclotron absorption (beyond the fluid description), which evidently will induce  other 
effects such as temperature anisotropy, heat flux, and beams formations in the velocity distribution 
function. Indeed, the spectrum of the compressible part of the velocity displayed in Fig. \ref{fig:spu} (middle panel)   for each of the simulations
  shows a rapid decay at $kd_i>1$ in the case of HVM, while HMHD and LF spectra
are  very similar and decay more slowly. 

   The different behavior between the compressible and solenoidal parts of the velocity is also conspicuous on the correlation of the out-of plane velocity $u_z$, which is 
  $49\%$ only between HMHD and HVM, while it reaches $83\%$ between LF and HVM. This is confirmed
  by inspection of the corresponding spectra displayed in Fig. \ref{fig:spu} (bottom panel). LF and
  HVM are rather close up to $kd_i \sim 1$, while the HMHD spectrum is quite different, even at $kd_i<1$. In order to understand this behavior, we performed an extra simulation within the bi-adiabatic approximation (i.e. with zero heat fluxes). It is interesting to note that the corresponding $u_z$ spectrum agrees with HVM up to roughly $kd_i=0.7$ (not shown) and then deviates significantly to catch up the HMHD spectrum. This indicates that at large scales, pressure anisotropy is a dominant correction, while at smaller scales, Landau damping plays a dominant role and contributes to reduce compressiblity \cite{hunana11}.
  
At the level of the  plasma density, correlations of $80\%$ and $92\%$ are
    respectively obtained in the cases HMHD/HVM and LF/HVM.
    The corresponding spectra, displayed in Fig. \ref{fig:sprho},  show a larger amplitude of the fluctuations for HVM at scales immediately smaller than $d_i$, a feature possibly related to the flattening of the density spectrum found in space plasma observations \cite{ale13} but not reproduced with fluid models.
   
   It is interesting at this stage to consider the ion distribution function.
 Figure \ref{fig:distrib} shows that it is significantly distorted in the $(v_x,v_z)$ -plane, with the presence of rings probably associated with ion-cyclotron absorption.
 This could be an explanation for the sharp decay of the compressible velocity spectrum that fluid models are unable to capture.


\subsection{Kinetic effects}
\label{kinetic}

\begin{figure*}
  \begin{center}
    \includegraphics[width=0.8\textwidth]{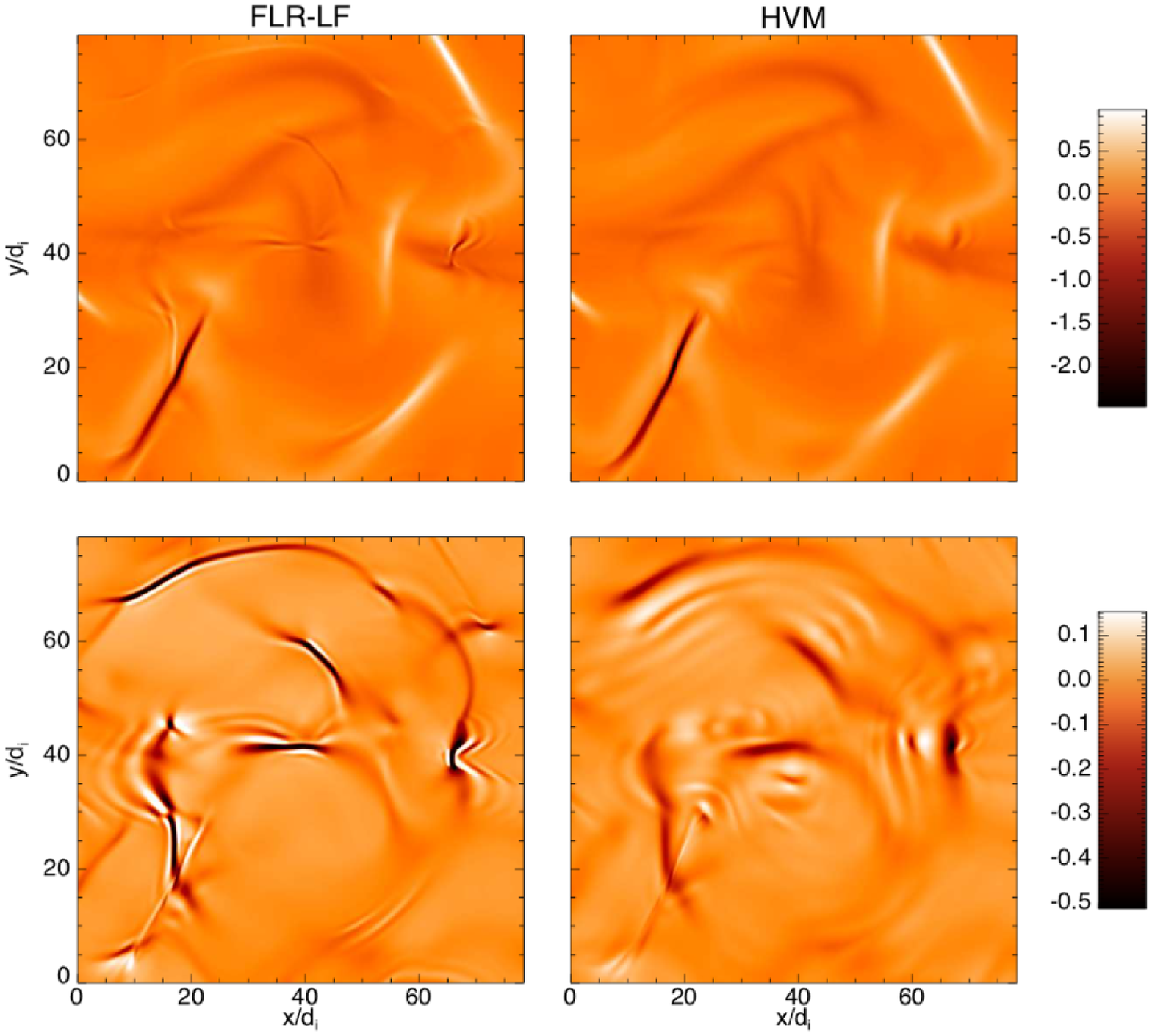}
  \end{center}
    \caption{Zoom of the out-of-plane current density $j_z$ (top) and proton velocity divergence $\xi$ (bottom) for the FLR-LF simulation with $\beta_e=0$ (left) and HVM (right) simulations at $t=25$. }
    \lab{fig:zoom}
\end{figure*}

In the last decades, several studies of turbulent collisionless plasmas, through {\it {in situ}} observations 
and numerical simulations, have shown that particle distribution functions are far from the typical configuration 
of thermodynamical equilibrium \cite[]{mar06,ser12,per13,per14a,per14b,val14,ser14a,ser14b,vas14,val16}.
Temperature anisotropy, particle heating and acceleration are observed in turbulence, localized in and 
near coherent structures. Thanks to kinetic approaches, and in particular to HVM simulations, 
we have the possibility to study in detail the non-Maxwellian effects on the ion velocity distribution.
On the other hand, the LF and FLR-LF models provide information on the low-frequency kinetic response
of the plasma in a turbulent context, missing nevertheless other kinetic effects such as the ion-cyclotron resonance.
In this part, we shall focus on temperature anisotropy and ion gyrotropic heat fluxes where a precise comparison can be performed between the LF and HVM simulations at $\beta_e=0$. An investigation of non-gyrotropic pressure tensor components will be performed in Section \ref{sec:FLR} where a FLR-LF run at $\beta_e=1$ will also be considered.

We start quantifying the non-Maxwellian features through the statistical analysis of the ion temperature
anisotropy, $T_{\perp}/T_{\parallel}$, defined as the ratio between the ion perpendicular and  parallel 
temperatures. The initial condition has been set up to have spatially
isotropic ion temperature. During the development of turbulence, the temperature does not remain isotropic
but presents local enhancements and depressions. Left panel of Fig.~\ref{fig:aniso} shows the PDFs of  $T_\perp/T_\|$, for both LF (blue solid-diamond line) and HVM (red solid-triangle line)
simulations at $t=60$. The behavior is similar: the PDFs elongate both in the parallel $(T_\perp/T_\|<1)$ and 
perpendicular $(T_\perp/T_\|>1)$ directions, displaying a strong anisotropic behavior, the anisotropy being more important 
along the perpendicular direction. 
Spearman correlation of the temperature ratio between the two simulations (see Table \ref{tb:spearman_kinetic}) is $92\%$, similar to the correlations
at the level of individual perpendicular ($91\%$) and parallel ($88\%$) temperatures.
 
Figure \ref{fig:tperpovertpar} displays a color plot of the temperature anisotropy $T_\perp/T_\|$ for the HVM run at $t=60$. It is  interesting to note that the regions of large (or small) values are not much correlated with the locations of the current sheets. The are rather related to regions of compression or dilation in the vicinity of these coherent structures and thus appear more diffuse (see also \cite{banon16,sorriso18} for a similar observation)
 
We now investigate the heat flux tensor, defined as the third order moment of the ion distribution function.
In particular, we look at the gyrotropic contribution characterized by the parallel $q_\|$ and perpendicular
$q_\perp$ heat fluxes. The PDFs
of ${q}_{\parallel}$ (solid line) and of ${q}_{\perp}$ (dashed line) for LF (in blue-diamond) and HVM (in red-triangle)
simulations are reported in the middle  panel of Fig.~\ref{fig:aniso}, at the maximum of the turbulent activity $(t=60)$. Although, 
${q}_{\parallel}$ and ${q}_{\perp}$ are both picked around zero, a preferential transfer of heat flux is
observed in the parallel direction. The LF and HVM results display similar distribution function. The agreement is better for 
the perpendicular heat flux, the PDF of the parallel heat flux being more populated for large deviation in the LF case. 
At the level of the Spearman
correlation coefficients, the tendency is however opposite, with $37\%$ and $79\%$ for $q_\perp$ and $q_\|$ respectively.
This difference in the quality of the correlations is very probably related to the hierarchy closure which, 
for technical reasons, involves a Pad\'e  approximant of the plasma response function
of higher order for the parallel than for the perpendicular quantities.

Figure~\ref{fig:aniso} (right panel) shows the time evolution, from $t=1$, of the Spearman correlation 
coefficient between LF and HVM simulations for the proton temperature anisotropy 
(red line), parallel heat flux (blue line) and velocity divergence 
(green line). After a brief transient in the case of the heat fluxes
originating from the low-frequency assumption underlying the hierarchy closure, a
decreasing behavior is recovered for all the quantities. 
In particular, both the runs start with an isotropic condition for ions, $T_\perp = T_\parallel$,
and the correlation is 1. Due to the development of the turbulence, the ion distribution
starts to deform, with generation of temperature anisotropy and heat flux, especially in 
the parallel direction. Both runs are able to describe this physics. However, due to the 
peculiarity of each description and to the limitation of the LF model, $C_s$ decreases
with the increase of turbulence. 
 
The most important discrepancy between the models appears in the compressibility.
Starting from $C_s(\xi) = 1$, the deviation becomes important very soon and, at $t=40$, 
the correlation is only  about $50\%$.

\section{The non-gyrotropic components of the pressure tensor} \label{sec:FLR}

The previous sections were concerned with a general description of the turbulence that develops in the 
HMHD, LF and HVM simulations, paying particular attention to the comparison between the fluid and kinetic modelings. 
Apart from a general description of velocity and magnetic structures together with their associated power spectra, 
kinetic features such as temperature anisotropy and heat fluxes have been discussed. 
The most important difference is related to the excess of small-scale kinetic compressibility in the fluid simulations 
as opposed to what is observed in the HVM simulation. 
It is now of interest to discuss the properties of the non-gyrotropic pressure tensor, comparing the outcome of HVM and  
FLR-LF simulations and also investigating how their amplitude relates to the current sheets.
Whereas for the LF case no numerical issue other than the necessity of adding  hyperdiffusivity to the density, 
pressure and heat flux equations was encountered, in the FLR-LF case it was not possible to continue the simulation 
past $t=30$ without having to use unreasonably large coefficients. 
There is probably not a unique reason for this numerical issue but one can mention that the high-order velocity derivatives 
present in the FLR terms are sensitive to sharp gradients. The Mach number of the $\beta_e=0$ simulation is large 
(the rms Mach number at $t=60$ reaches $0.39$), permitting the development of rather strong shocks in the FLR-LF run, 
as revealed by the sharp dark lines in the plot of the velocity divergence at $t=25$ (see left bottom panel of Fig. \ref{fig:zoom}). 
These shocks do not preclude to continue the simulation. 
The numerical difficulty is rather related to the destabilization of a current sheet after it has interacted with a traveling shock. 
Bottom left corner of the panels  for $\nabla\cdot\boldsymbol{u}$ and $j_z$ in Fig. \ref{fig:zoom} (left panel) 
shows such an interaction happening for the FLR-LF run. In the HVM simulation, these shocks are mollified and hardly 
visible at the level of the current sheet (right panels of Fig. \ref{fig:zoom}). Other smooth oscillations are nevertheless 
present in the Vlasov case. This is consistent with the previous remark concerning the difference between the HVM and the 
LF simulation at the level of the compressible part of the velocity field.

\begin{figure}
  \begin{center}
    \includegraphics [width=0.48\textwidth]{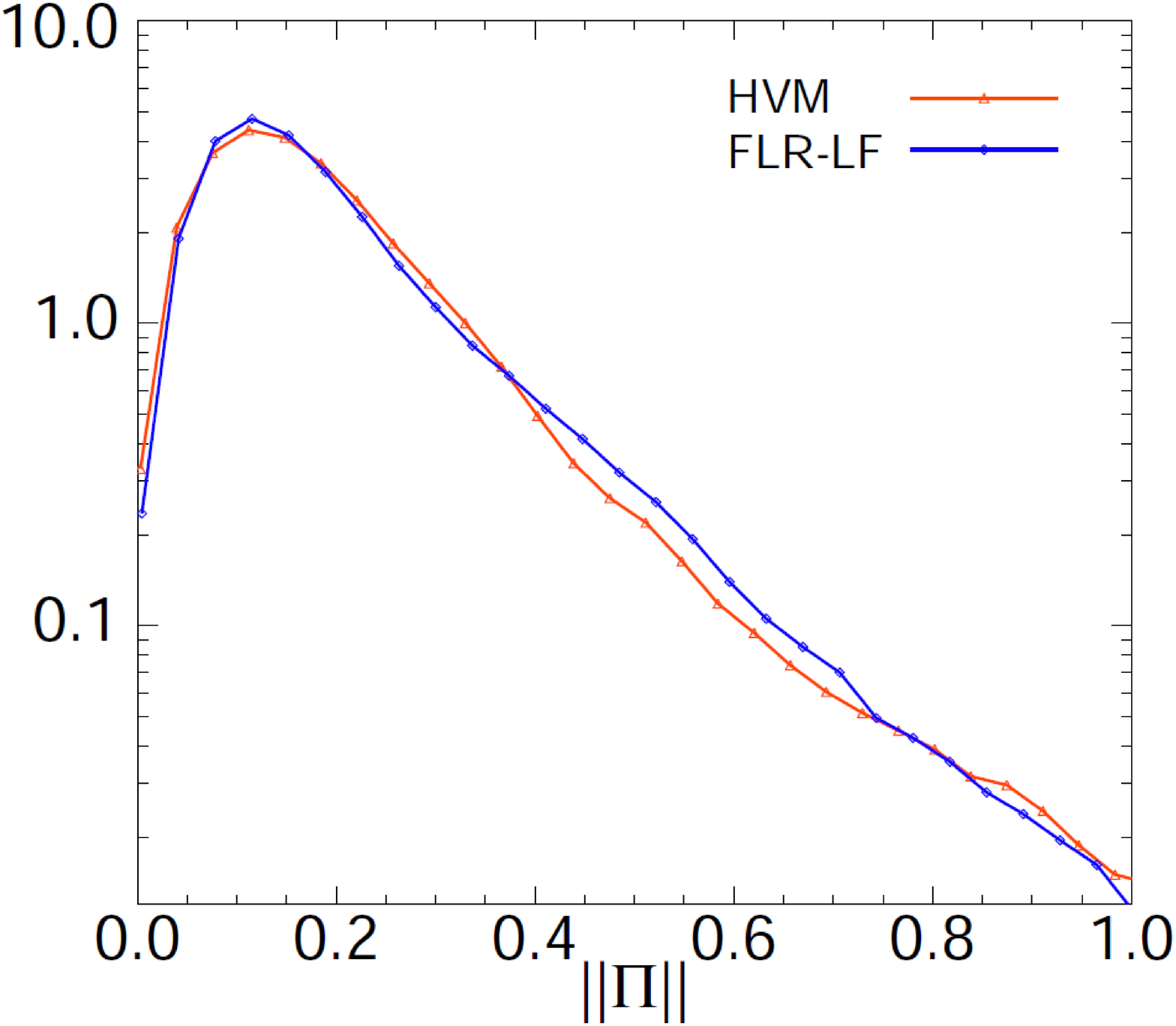}
    \caption{PDFs of the 
      norm of the non-gyrotropic tensor, $||{\mathbf \Pi}||$ for FLR-LF (blue solid-diamond line) 
      and HVM (red solid-triangle line) simulations at $t=60$.}
    \label{fig:histopinorm}
  \end{center}
\end{figure}

\begin{figure}
  \begin{center}
    \includegraphics [width=0.48\textwidth]{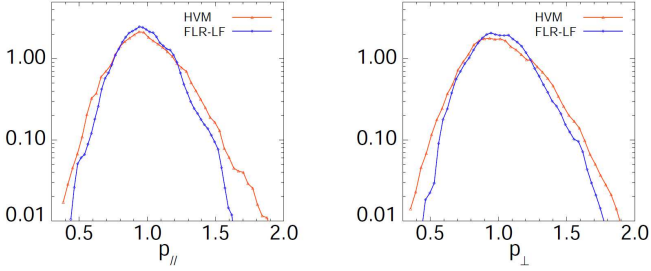}
  \end{center}
  \caption{PDFs of gyrotropic pressure components, $p_{\|}$ and $p_{\perp}$ (left and right), for FLR-LF (blue solid-diamond line) 
    and HVM (red solid-triangle line) simulations at $t=60$.}
  \label{fig:pgyro}
\end{figure}

\begin{figure}
  \begin{center}
    \includegraphics [width=0.48\textwidth]{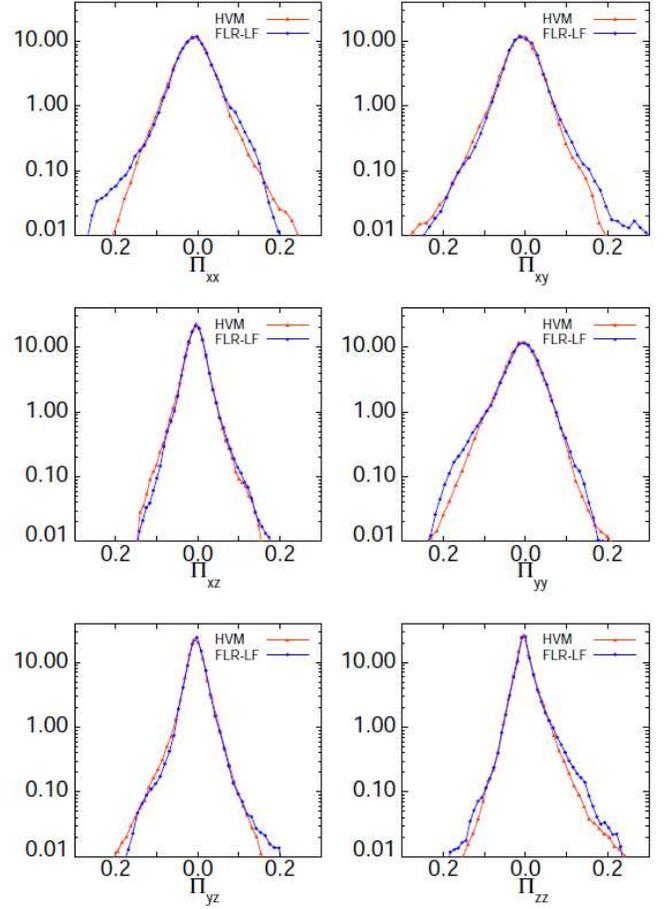}
  \end{center}
  \caption{PDFs of the various non-gyrotropic tensor components, ${\mathbf \Pi}_{xx}$ and ${\mathbf \Pi}_{xy}$ (top left and right), ${\mathbf \Pi}_{xz}$ and ${\mathbf \Pi}_{yy}$ (middle left and right) and 
    ${\mathbf \Pi}_{yz}$ and ${\mathbf \Pi}_{zz}$ (bottom left and right) for FLR-LF (blue solid-diamond line) 
    and HVM (red solid-triangle line) simulations at $t=60$.}
  \label{fig:Picomponents}
\end{figure}

\begin{figure}
  \begin{center}
    \includegraphics [width=0.48\textwidth]{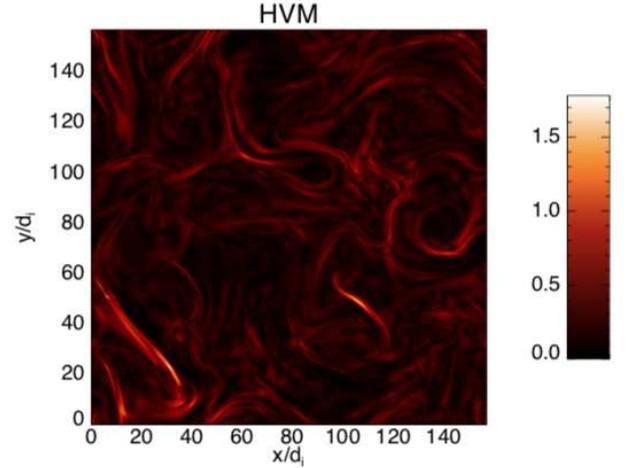}
  \end{center}
  \caption{Color scale plot of the norm of the non-gyrotropic pressure tensor  $||{\mathbf \Pi}||$ for the HVM run at $t=60$}
  \label{fig:pinorm}
\end{figure}

\subsection{Comparison with a FLR-LF simulation at $\beta_e=1$}

As will be seen, the FLR pressure tensor is weakly sensitive to the value of the electron beta parameter 
(at least as long as no strong shocks form), justifying a comparison between the HVM simulation (at $\beta_e=0$) 
and a FLR-LF simulation performed at $\beta_e=1$. Indeed, due to a lower value of the Mach number, 
a FLR-LF run at $\beta_e=1$ can be completed without too much difficulty, but nevertheless at the expense of using 
larger hyperviscosity coefficients (note also that a comparison with a HVM at $\beta_e=1$ would be interesting 
but not presently possible due to computational power limitations). This comparison also contributes to a validation 
of the FLR formulation used in the FLR-LF model.

Before displaying the FLR pressure tensor components, let us recall some definitions.
The pressure tensor can be written as
\beq
{\bf p} = {\bf p^G} + {\bf \Pi} \ ,
\eeq 
where ${\bf p^G}$ the gyrotropic pressure tensor (built from the parallel $p_\|$
and perpendicular $p_\perp$ pressures)
and ${\bf \Pi}$ is due to the FLR effects. The pressure tensor, together with its gyrotropic and nongyrotropic
components are symmetric, with the additional condition ${\rm tr}({\bf \Pi}) = 0$. Note that all the FLR pressure tensor components are calculated in a basis using the direction of the local magnetic field.

Figure~\ref{fig:histopinorm} shows the PDFs of the norm of ${\bf \Pi}$ for both FLR-LF (with $\beta_e=1$; blue solid-diamond line) and HVM (with $\beta_e=0$; red solid-triangle line) simulations at $t=60$. 
The norm is defined as 
\beq
|| {\bf \Pi} || = \sum_{i,j} | {\bf \Pi}_{ij} | \ ,
\eeq
where the sum includes all components. Both simulations show that the contribution of 
the non-gyrotropic term to the total pressure is not negligible and that the
  distribution functions are very similar. 
This surprising result reveals that the FLR pressure tensor is not very sensitive to the Mach number (in particular the compressible component of the velocity). The PDF of the gyrotropic  components is however more directly influenced by a difference in the $\beta_e$ parameter of the two simulations, as demonstrated in Fig. \ref{fig:pgyro} showing a greater deviation between the two runs.

 Although there is no point in making more precise comparisons between the two runs which use different values of the $\beta_e$ parameter, let us mention that the Spearman correlation FLR-LF/HVM for $|| {\bf \Pi} ||$ 
	is  $49\%$.

When considering the PDFs of the individual
  cartesian components of the non-gyrotropic pressure tensor, we note that the components involving a $z$ index are more peaked (see Fig. \ref{fig:Picomponents}). As previously noted for the norm, a surprisingly good agreement is found between the FLR-LF and the HVM simulations, demonstrating that the fluid modeling captures the essential properties of the non-gyrotropic pressure tensor.
  
Moreover, a dependence of $|| {\bf \Pi} ||$ on the level of turbulence has been observed (not shown here). Indeed, it increases 
when the magnitude of the turbulent fluctuations increases in the system, reaching its maximum when turbulent activity is 
maximum ($t=60$).  Figure \ref{fig:pinorm} shows a color plot of $|| {\bf \Pi} ||$ from  the HVM run at $t=60$. It indeed seems 
rather well correlated with the out-of-plane current although displaying multiple filamentary structures where the current only shows isolated sheets. A very similar picture is obtained with the FLR-LF run (not shown).

\section{Conclusion}
\label{conclusion}

Two-dimensional simulations of turbulence development in a collisionless plasma subject to an out-of-plane ambient
magnetic field were performed using HVM  and fluid models
such as HMHD, LF and FLR-LF. Both in HVM and LF, electrons are cold, but an extra FLR-LF simulation with $\beta_e=1$ is used to illustrate the comparison of the non-gyrotropic pressure tensor between FLR-LF and HVM.
In all the cases, initial conditions are large-scale, moderate-amplitude perturbations of a homogeneous equilibrium state.
A main observation concerns the capabilities of fluid models to approach HVM simulations, depending on the
considered fields. The magnetic field is globally well described, especially the in-plane component, up to wavenumbers
such that $kd_i \approx 5$. For the velocity field, the agreement stops at $kd_i=1$ for the in-plane components. The out-of-plane velocity is poorly described
at all  scales by HMHD, but satisfactorily reproduced by LF up to $kd_i \approx 1$. We checked that Landau damping
contributes to this agreement. Interestingly, the  compressible (in-plane) velocities of HMHD and LF are very similar 
and correctly reproduce HVM up to $kd_i \approx 1$. At this wavenumber, the HVM energy spectrum of this component displays
a sharp break. Fast kinetic effects, such as ion-cyclotron resonance, not retained in fluid approaches, could be at the origin of this effect.
Concerning quantities not retained by HMHD, a satisfactory agreement is obtained
between LF and HVM at the level of temperature anisotropy and heat fluxes. A significant temperature anisotropy is generated, with a bias towards the perpendicular component, the more intense fluctuations being rather spread out and located in a broad vicinity of current sheets.
The FLR-LF run at $\beta_e=1$ displays a surprisingly good agreement with HVM regarding non-gyrotropic pressure components, showing their relative insensitivity to the $\beta_e$ parameter. The fluctuation amplitude of these components almost reaches $50\%$ of the gyrotropic ones, with the most intense regions located near current sheets, showing
the necessity of taking them into account in a small-scale modeling of the plasma. 

Further developments
should include comparisons in three space dimensions where linear waves, not present in two dimensions but accurately
described by FLR-LF, are expected to significantly affect the dynamics.
\bigskip

\acknowledgments
DP, SS and IZ acknowledge support from the Faculty of the European Space Astronomy Centre (ESAC).
Part of this work has been supported by the Agenzia Spaziale Italiana under the Contract No. ASI-INAF 2015-039-R.O `Missione M4 di ESA: Partecipazione Italiana alla fase di assessment della missione THOR'. The HVM simulations have been performed on the Newton parallel machine at the Department of Physics at University of Calabria (Italy). 
This work was also granted access to the HPC resources of CINES/IDRIS
under the allocation 2016047042.  Part of the fluid computations have been
done on the 'Mesocentre SIGAMM' machine, hosted by Observatoire de
la C\^ote d'Azur.



\begin{thebibliography}{}

\bibitem[Kolmogorov(1941)]{kol41}
A. N. Kolmogorov,
Dokl. Akad. Nauk SSSR {\bf 30}, 9 (1941).

\bibitem[Bruno \& Carbone(2013)]{bru13}
R. Bruno, and V. Carbone, 
Living Rev. Solar Phys. {\bf 10}, 2 (2013).

\bibitem[Belcher \& Davis(1971)]{bel71}
J. W. Belcher, and L. Davis, 
J. Geophys. Res. {\bf 76}, 3534 (1971).

\bibitem[Matthaeus \& Goldstein(1982)]{mat82}
W. H. Matthaeus, and M. L. Goldstein, 
J. Geophys. Res. {\bf 87}, 6011 (1982).

\bibitem[Tu \& Marsch(1995)]{tu95}
C. Y. Tu, and E. Marsch, 
Space Sci. Rev. {\bf 73}, 1 (1995).

\bibitem[Sahraoui et  al.(2010)]{sahraoui10}
F. Sahraoui, M. L. Goldstein, G. Belmont, P. Canu, and L. Rezeau,
Phys. Rev. Lett. {\bf 105}, 131101 (2010).

\bibitem[Leamon et al.(1998)]{lea98}
R. J. Leamon, C. W. Smith, N. F. Ness, W. H. Matthaeus, and H. K. Wong, 
J. Geophys. Res. {\bf 103}, 4775 (1998).

\bibitem[Leamon et al.(2000)]{lea00}
R. J. Leamon, W. H. Matthaeus, C. W. Smith, G. P. Zank, D. J. Mullan, and S. Oughton,
Astrophys. J. {\bf 537}, 1054 (2000).

\bibitem[Bale et al.(2005)]{bal05}
S. D. Bale, P. J. Kellogg, F. S. Mozer, T. S. Horbury, and H. Reme, 
Phys. Rev. Lett. {\bf 94}, 215002 (2005).

\bibitem[Smith et al.(2006)]{smi06}
C. W. Smith, K. Hamilton, and B. J. Vasquez, 
Astrophys. J. Lett. {\bf 645}, L85 (2006).

\bibitem[Bourouaine et al.(2012)]{bou12}
S. Bourouaine, O. Alexandrova, E. Marsch, and M. Maksimovic, 
Astrophys. J. {\bf 749}, 102 (2012).

\bibitem[Alexandrova et al.(2012)]{alexandrova12}
O. Alexandrova, V. Carbone, P. Veltri, and L. Sorriso-Valvo, 
Astrophys. J. Lett. {\bf 674}, 1153 (2012).

\bibitem[Saharaoui et al.(2013)]{sahraoui13}
 F. Sahraoui, S. Y. Huang, G. Belmont, M. L. Goldstein, A. Retin\`o, P. Robert, and J. De~Patoul, 
Astrophys. J. {\bf 777}, 15 (2013).

\bibitem[Matthaeus et al.(2014)]{bill14}
W. H. Matthaeus, S. Oughton, K. T. Osman, S. Servidio, M. Wan, S. P. Gary, M. A. Shay, F. Valentini, V. Roytershteyn, and H. Karimabadi,
Astrophys. J. {\bf 790}, 155 (2014).

\bibitem[Alexandrova et al.(2007)]{ale07}
O. Alexandrova, V. Carbone, P. Veltri, and L. Sorriso-Valvo, 
Planet. Spa. Sci. {\bf 55}, 2224 (2007).

\bibitem[Alexandrova et al.(2008)]{ale08}
O. Alexandrova, V. Carbone, P. Veltri, and L. Sorriso-Valvo, 
Astrophys. J. {\bf 674}, 1153 (2008).

\bibitem[Hamilton et al.(2008)]{ham08}
K. Hamilton, C. W. Smith, B. J. Vasquez, and R. J. Leamon, 
J. Geophys. Res. {\bf 113}, A01106 (2008).

\bibitem[Salem et al.(2012)]{sal12}
C. S. Salem, G. G. Howes, D. Sundkvist, S. D. Bale, C. C. Chaston, C. H. K. Chen, and F. S. Mozer, 
Astrophys. J. Lett. {\bf 745}, L9 (2012).

\bibitem[Kiyani et al.(2013)]{kiy13}
K. H. Kiyani, S. C. Chapman, F. Sahraoui, B. Hnat, O. Fauvarque, and Yu. V. Khotyaintsev, 
Astrophys. J. {\bf 763}, 10 (2013).

\bibitem[Perrone et al.(2016)]{per16}
D. Perrone, O. Alexandrova, A. Mangeney, M. Maksimovic, C. Lacombe, V. Rakoto, J. C. Kasper, and D. Jovanovic, 
Astrophys. J. {\bf 826}, 196 (2016).

\bibitem[Perrone et al.(2017)]{per17}
D. Perrone, O. Alexandrova, O. W. Roberts, S. Lion, C. Lacombe, A. Walsh, M. Maksimovic, and I. Zouganelis, 
Astrophys. J. {\bf 849}, 49 (2017).

\bibitem[Marsch(2006)]{mar06}
E. Marsch, 
Living Rev. Solar Phys. {\bf 3}, 1 (2006).

\bibitem[Hollweg \& Isenberg(2002)]{hol02}
J. V. Hollweg, and P. A. Isenberg, 
J. Geophys. Res. {\bf 107}, 1147 (2002).

\bibitem[Valentini et al.(2005)]{val05pre}
F. Valentini, V. Carbone, P. Veltri, and A. Mangeney,
Phys. Rev. E {\bf 71}, 017402 (2005).

\bibitem[Schekochihin et al.(2009)]{sch09}
A. A. Schekochihin, S. C. Cowley, W. Dorland, G. W. Hammett, G. G. Howes, E. Quataert, and T. Tatsuno,
Astrophys. J. {\bf 182}, 310 (2009).

\bibitem[Retin\`o et al.(2007)]{ret07}
A. Retin\`o, D. Sundkvist, A. Vaivads, F. Mozer, M. Andr\'e, and C. J. Owen, 
Nature Phys. {\bf 3}, 236 (2007).

\bibitem[Sundkvist et al.(2007)]{sun07}
S. Sundkvist, A. Retin\`o, A. Vaivads, and S. D. Bale, 
Phys. Rev. Lett. {\bf 99}, 025004 (2007).

\bibitem[Osman et al.(2011)]{osm11}
K. T. Osman, W. H. Matthaeus, A. Greco, and S. Servidio,  
Astrophys. J. Lett. {\bf 727}, L11 (2011).

\bibitem[Chasapis et al.(2015)]{cha15} 
A. Chasapis, A. Retin\`o, F. Sahraoui, A. Vaivads, Yu. V. Khotyaintsev, D. Sundkvist, A. Greco, L. Sorriso-Valvo, and P. Canu, 
Astrophys. J. Lett. {\bf 804}, L1 (2015).

\bibitem[Chasapis et al.(2017)]{cha17} 
A. Chasapis, W. H. Matthaeus, T. N. Parashar, O. LeContel, A. Retin\`o, H. Breuillard, Yu. V. Khotyaintsev, A. Vaivads, B. Lavraud, E. Eriksson,
Astrophys. J. {\bf 836}, 247 (2017).

\bibitem[Pezzi et al.(2016)]{pezzi16}
O. Pezzi, F. Valentini, and P. Veltri, 
Phys. Rev. Lett. {\bf 116}, 145001 (2016).

\bibitem[Pezzi (2017)]{pezzi17}
O. Pezzi, 
J. Plasma Phys. {\bf 83}, 555830301 (2017).

\bibitem[Chandran et al.(2010)]{cha10}
B. D. G. Chandran, B. Li, B. N. Rogers, E. Quataert, and K. Germascherwski, 
Astrophys. J. {\bf 720}, 503 (2010).



\bibitem[Gurnett \& Bhattacharjee(2005)]{gur05}
D. A. Gurnett, and A. Bhattacharjee, 
{\it Introduction to Plasma Physics: With Space and Laboratory Applications},
Cambridge University Press (2005).

\bibitem[Zank \& Matthaeus(1992)]{zan92}
G. P. Zank, and W. H. Matthaeus, 
J. Plasma Phys. {\bf 48}, 85 (1992).

\bibitem[Ghosh \& Goldstein(1997)]{gho97}
S. Ghosh, and M. L. Goldstein, 
J. Plasma Phys. {\bf 57}, 129 (1997).

\bibitem[Kingsep et al.(1987)]{kin87}
A. S. Kingsep, K. V. Chukbar, and V. V. Ian'kov, 
Voprosy Teorii Plazmy {\bf 16}, 209 (1987). 

\bibitem[Gordeev et al.(1994)]{gor94}
A. V. Gordeev, A. S. Kingsep, and L. I. Rudakov, 
Phys. Rep. {\bf 243}, 215 (1994).

\bibitem[Cho \& Lazarian(2004)]{cho04}
J. Cho, and A. Lazarian, 
Astrophys. J. Lett. {\bf 615}, L41 (2004).

\bibitem[Snyder et al.(1997)]{SHD97}
P. B. Snyder, G. W. Hammett, and W. Dorland, 
Phys. Plasmas {\bf 4}, 3974 (1997).

\bibitem[Sulem \& Passot(2015)]{sul15}
P. L. Sulem, and T. Passot, 
J. Plasma Phys. {\bf 81}, 325810103 (2015).

\bibitem[Brizard \& Hahm(2007)]{bri07}
A. J. Brizard, and T. S. Hahm,  
Rev. Mod. Phys. {\bf 79}, 421 (2007).

\bibitem[Howes et al.(2006)]{Howes06}
  G. G. Howes, S. C. Cowley, W. Dorland, G. W. Hammett, E. Quataert, and A. A. Schekochihin,
  Astrophys. J. {\bf 651}, 590 (2006).

\bibitem[Numata et al.(2010)]{Numata10}
  R. Numata, G. G. Howes, T. Tatsono, M. Barnes, and W. Dorland, 
  J. Comp. Phys. {\bf 229}, 9347 (2010).

\bibitem[Birdsall \& Langdon(1985)]{bir85}
C. K. Birdsall, and A. B. Langdon, 
{\it Plasma Physics via Computer Simulation},
McGraw-Hill Book Company, Singapore (1985).

\bibitem[Parashar et al.(2009)]{par09}
T. N. Parashar, M. A. Shay, P. A. Cassak, and W. H. Matthaeus,  
Phys. Plasmas {\bf 16}, 032310 (2009).

\bibitem[Karimabadi et al.(2013)]{kar13}
H. Karimabadi, V. Roytershteyn, M. Wan, W. H. Matthaeus, W. Daughton, P. Wu, M. Shay, B. Loring, J. Borovsky, E. Leonardis, et al., 
Phys. Plasmas {\bf 20}, 012303 (2013).

\bibitem[Mangeney et al.(2002)]{man02}
A. Mangeney, F. Califano, C. Cavazzoni, and P. Tr\'avn\'icek,  
J. Comp. Phys. {\bf 179}, 495 (2002).

\bibitem[Valentini et al.(2005)]{val05}
F. Valentini, P. Veltri, and A. Mangeney,  
J. Comp. Phys. {\bf 210}, 730 (2005).

\bibitem[Valentini et al.(2007)]{val07}
F. Valentini, P. Tr\'avn\'icek, F. Califano, P. Hellinger, and A. Mangeney,  
J. Comp. Phys. {\bf 225}, 753 (2007).

\bibitem[Valentini et al.(2008)]{val08}
F. Valentini, P. Veltri, F. Califano, and A. Mangeney,  
Phys. Rev. Lett. {\bf 101}, 025006 (2008).

\bibitem[Valentini \& Veltri(2009)]{val09}
F. Valentini, and P. Veltri,  
Phys. Rev. Lett. {\bf 102}, 225001 (2009).

\bibitem[Valentini et al.(2010)]{val10}
F. Valentini, F. Califano, and P. Veltri,  
Phys. Rev. Lett. {\bf 104}, 205002 (2010).

\bibitem[Valentini et al.(2011)]{val11}
F. Valentini, D. Perrone, and P. Veltri,  
Astrophys. J. {\bf 739}, 54 (2011).

\bibitem[Perrone et al.(2011)]{per11}
D. Perrone, F. Valentini, and P. Veltri,  
Astrophys. J. {\bf 741}, 43 (2011).

\bibitem[Servidio et al.(2012)]{ser12}
S. Servidio, F. Valentini, F. Califano, and P. Veltri,  
Phys. Rev. Lett. {\bf 108}, 045001 (2012).

\bibitem[Perrone et al.(2013)]{per13}
D. Perrone, F. Valentini, S. Servidio, S. Dalena, and P. Veltri,  
Astrophys. J. {\bf 762}, 99 (2013).

\bibitem[Perrone et al.(2014a)]{per14a}
D. Perrone, S. Bourouaine, F. Valentini, E. Marsch, and P. Veltri,  
J. Geophys. Res. {\bf 119}, 2400 (2014).

\bibitem[Perrone et al.(2014b)]{per14b}
D. Perrone, F. Valentini, S. Servidio, S. Dalena, and P. Veltri,  
Eur. Phys. J. D {\bf 68}, 209 (2014).

\bibitem[Valentini et al.(2014)]{val14}
F. Valentini, S. Servidio, D. Perrone, F. Califano, W. H. Matthaeus, and P. Veltri, 
Phys. Plasmas {\bf 21}, 082307 (2014).

\bibitem[Servidio et al.(2014a)]{ser14a}
S. Servidio, K. T. Osman, F. Valentini, D. Perrone, F. Califano, S. Chapman,W. H. Matthaeus, and P. Veltri, 
Astrophys. J. Lett. {\bf 781}, L27 (2014).

\bibitem[Servidio et al.(2015)]{ser14b}
S. Servidio, F. Valentini, D. Perrone, A. Greco, F. Califano, W. H. Matthaeus, and P. Veltri, 
J. Plasma Phys. {\bf 81}, 325810107 (2015).

\bibitem[V\'asconez et al.(2014)]{vas14}
C. L. V\'asconez, F. Valentini, E. Camporeale, and P. Veltri,  
Phys. Plasmas {\bf 21}, 112107 (2014).

\bibitem[Valentini et al.(2016)]{val16}
F. Valentini, D. Perrone, S. Stabile, O. Pezzi, S. Servidio, R. De Marco, F. Marcucci, R. Bruno, B. Lavraud, J. De Keyser, et al.,  
New J. Phys. {\bf 18}, 125001 (2016).

\bibitem[Pezzi et al.(2017a)]{pezzi17a} 
O. Pezzi, T. N. Parashar, S. Servidio, F. Valentini, C. L. V\'asconez, Y. Yang, F. Malara, W. H. Matthaeus, and P. Veltri,
Astrophys. J. {\bf 834}, 166 (2017).

\bibitem[Pezzi et al.(2017b)]{pezzi17b} 
O. Pezzi, T. N. Parashar, S. Servidio, F. Valentini, C. L. V\'asconez, Y. Yang, F. Malara, W. H. Matthaeus, and P. Veltri,
J. Plasma Phys., {\bf 83}, 90583015 (2017).

\bibitem[Henri et al.(2013)]{hen13}
P. Henri, S. S. Cerri, F. Califano, F. Pegoraro, C. Rossi, M. Faganello, O. Sebek, P. M. Tr\'avn\'icek, P. Hellinger, J. T. Frederiksen, et al.,  
Phys. Plasmas {\bf 20}, 102118 (2013).

\bibitem[Parashar et al.(2015)]{par15}
T. N. Parashar, C. Salem, R. T. Wicks, H. Karimabadi, S. P. Gary, and W. H. Matthaeus,  
J. Plasma Phys. {\bf 81}, 905810513 (2015).

\bibitem[Passot et al.(2014)]{pas14}
T. Passot, P. Henri, D. Laveder, and P. L. Sulem,  
Eur. Phys. J. D {\bf 68}, 207 (2014).

\bibitem[Hunana et al.(2013)]{hun13}
P. Hunana, M. L. Goldstein, T. Passot, P. L. Sulem, D. Laveder, and G. P. Zank,
Astrophys. J. {\bf 766}, 93 (2013).

\bibitem[Kobayashi et al. (2017)]{koba17}
S. Kobayashi, F. Sahraoui, T. Passot, D. Laveder, P. L. Sulem, S. Y. Huang, P. Henri, and R. Smets, 
Astrophys. J. {\bf 839}, 122 (2017).

\bibitem[Cheng \& Knorr(1976)]{che76}
C. Z. Cheng, and G. Knorr,  
J. Comp. Phys. {\bf 22}, 330 (1976).

\bibitem[Matthews(1994)]{mat94}
A. P. Matthews, 
J. Comput. Phys. {\bf 112}, 102 (1994). 

\bibitem[Peyret \& Taylor(1986)]{pey86}
R. Peyret, and T. D. Taylor, 
{\it Computational Methods for Fluid Flow},
New York, Springer (1986).

\bibitem[Mininni \& Pouquet(2009)]{min09}
P. D. Mininni, and A. Pouquet, 
Phys. Rev. E {\bf 80}, 025401 (2009).

\bibitem[Hunana et al. (2011)]{hunana11}
P. Hunana, D. Laveder, T. Passot, P. L. Sulem, and D. Borgogno, 
Astrophys. J. {\bf 743}, 128 (2011).

\bibitem[Alexandrova et al.(2013)]{ale13}
O. Alexandrova, C. H. K. Chen, L. Sorriso-Valvo, T. S. Horbury, and S. D. Bale, 
Space Sci. Rev. {\bf 178}, 101 (2013).

\bibitem[Ba\~n\'on Navarro et al.(2016)]{banon16}
A. Ba\~n\'on Navarro, B. Teaca, D. Told, D. Groselj, P. Crandall, and F. Jenko, 
Phys. Rev. Lett. {\bf 117}, 245101 (2016).

\bibitem[Sorriso et al.(2018, under review)]{sorriso18}
L. Sorriso-Valvo, D. Perrone, O. Pezzi, F. Valentini, S. Servidio, I. Zouganelis, and P. Veltri,  
under review on J. Plasma Phys.

\end{thebibliography}
\end{document}